\documentclass{article}
\usepackage{hiph-art}
\volnumber{??} \issuenumber{?} \edyear{2006}                             
\frompage{000} \topage{000}                                              
\recrevdate{1 July 2006}                                              
\def\rightmark{Kaon production from 1 to 40 A GeV}
\def\leftmark{A.B. Larionov, U. Mosel, and M. Wagner}
 
\title{Kaon production from 1 to 40 A GeV}
\authors{ 
{A.B. Larionov$^{1,2}$, U. Mosel$^1$, and M. Wagner$^1$ %
\index{Larionov, A.B.} 
\index{Mosel, U.} 
\index{Wagner, M.}
}\\[2.812mm]
{\normalsize
\hspace*{-8pt}$^1$ Institut f\"ur Theoretische Physik, Universit\"at Giessen,\\ 
D-35392 Giessen, Germany\\[0.2ex]
\hspace*{-8pt}$^2$ RRC "I.V. Kurchatov Institute",\\ 
123182 Moscow, Russia
}} 
 
\abstract{
Kaon production is studied within the Giessen Boltzmann-Uehling-Uhlenbeck 
(GiBUU) model. Results are compared with experiment and with other models.
The influence of the kaon potential on the kaon azimuthal distributions
at SIS energies is considered. We also discuss the role of the many-body 
collisions at high-density phase of reaction.}


\makeindex
\begin{document}
 
\maketitle

\section{ Motivation }

Strangeness production has been a hot topic of heavy-ion studies for about 
20 years. Due to strangeness conservation, kaons can not be absorbed in 
the nuclear medium. The kaon-nucleon scattering cross section is also quite 
small ($\sim 10$ mb). Thus, kaons deliver a signal from the high-density 
phase of a reaction. This property of kaons was extremely useful to extract 
information on the nuclear equation of state from kaon multiplicities 
at 1-2 A GeV \cite{Sturm01,Fuchs01}. Another interesting observable at SIS 
energies is the kaon collective flow which is sensitive to the kaon 
potential. The measurements of the K$^+$ in-plane flow have been performed 
by the FOPI Collaboration \cite{Herr99,Crochet00}.
The analysis within the T\"ubingen QMD model \cite{Zheng04} has revealed
that the data \cite{Herr99} can be described only by using a strongly
repulsive K$^+$ potential given by the Brown-Rho (BR) parametrization
($U_K(\rho_0) \simeq 30$ MeV). Recently the KaoS Collaboration published
data on K$^+$ azimuthal distributions \cite{Uhlig05} which were 
analysed in \cite{LM05}.

The strange particle and pion multiplicities have also been measured
at higher energies: AGS \cite{Ahle00,Ahle00_1} and SPS \cite{Afan02,Friese04}.
The most interesting observable is the $K^+/\pi^+$ ratio plotted vs the
beam energy, which has a maximum at $E_{lab} \sim 20$ A GeV 
(c.f. Fig.~\ref{fig:ratio} below). So far, the transport models based on
hadronic and string degrees of freedom have failed to describe the maximum,
which could also be a manifestation of the transition to the quark-gluon 
plasma phase. 

We present here some selected results of calculations within the GiBUU model 
on the kaon azimuthal distributions at SIS energies and on
the  $K^+/\pi^+$ ratio and the slopes of  kaon $m_\perp$-spectra at AGS-SPS
enegies. A full analysis can be found in \cite{LM05,WLM05}.

A brief description of the model is given in Sect. 2. Sect. 3 contains 
numerical results. Sect. 4 summarises our study.  

\section{GiBUU model}

Our calculations are based on the GiBUU model in version of Refs. 
\cite{EBM99,MEPhD}. The model describes a heavy-ion collision explicitly 
in time as a sequence of elementary two-particle collisions and resonance 
decays. Between
collisions, the particles either propagate in the mean field (optionally)
or along straight trajectories. The baryon-baryon
collisions at $\sqrt{s}<2.6$ GeV and the meson-baryon collisions at
$\sqrt{s}<2$ GeV are treated within the resonance model. At larger $\sqrt{s}$,
the string model is applied (c.f. Ref. \cite{Falter04} for details).

At SIS energies, we treat strangeness production perturbatively and use the 
mean field potentials for the propagation of baryons and kaons \cite{com}.
For the baryons, the soft momentum-dependent mean field is used
(SM, K=215 MeV, see \cite{MEPhD} for details), which is well suited 
to reproduce the nucleon collective flows \cite{LCGM00} and kaon
multiplicities in Au+Au and C+C systems \cite{Fuchs01,LM05}.

The $K^\pm$ single-particle energies contain both vector and scalar parts
(c.f. \cite{Zheng04,BR96}):
\begin{equation}
   \omega^\pm({\bf k}) = \pm V_0 + \sqrt{{\bf k}^{*2} + m_K^{*2}}~, \label{omega}
\end{equation}
where ${\bf k}^* \equiv {\bf k} \mp {\bf V}$ is the kaon kinetic momentum,
\begin{equation} 
   V^\mu = \frac{3}{8f_\pi^{*2}} j_B^\mu
\end{equation} 
is the kaon vector potential, $j_B^\mu = <\bar B \gamma^\mu B>$ is
the baryon four-current.
\begin{equation}
   m_K^* = \sqrt{ m_K^2 - \frac{\Sigma_{KN}}{f_\pi^2} \rho_s + V_\mu V^\mu }
\end{equation}
is the kaon effective (Dirac) mass, where $\rho_s = <\bar B B>$
is the baryon scalar density, $f_\pi = 0.093$ GeV is the vacuum pion decay
constant, $m_K=0.496$ GeV is the bare kaon mass. $\Sigma_{KN}$ is the
kaon-nucleon sigma term.

Following Ref. \cite{Zheng04}, we will use the BR \cite{BR96} and the Ko-Li 
(KL) \cite{LK95} parametrizations of the kaon single-particle energy which 
differ by the choice of the kaon-nucleon sigma-term and of the in-medium pion
decay constant $f_\pi^*$. In case of BR (KL) $\Sigma_{KN}=0.450~(0.350)$ GeV
and $(f_\pi^*/f_\pi)^2=0.6~(1)$.

\begin{figure}[htb]
\vspace*{-0.5cm}
                  \insertplot{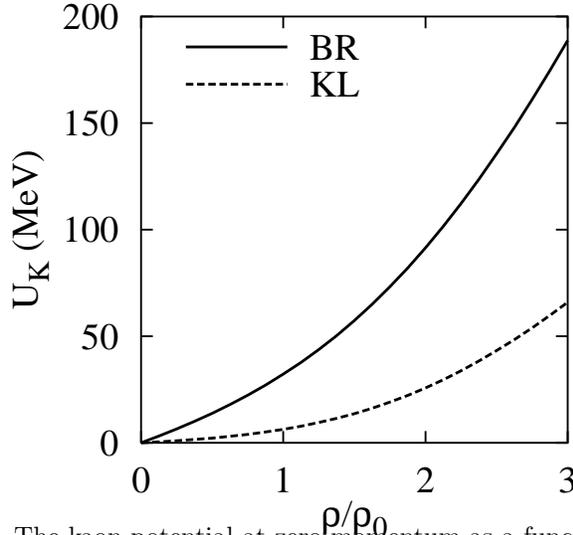}
\vspace*{-1.0cm}
\caption[]{The kaon potential
at zero momentum as a function of the nuclear matter density.}
\label{fig:ukaon}
\end{figure}

The $K^+$ potential 
$U_K({\bf k}) = \omega^+_K({\bf k}) -  \sqrt{{\bf k}^2 + m_K^2}$
at zero momentum is depicted in Fig.~\ref{fig:ukaon} as a function of the 
baryon density. The BR potential is much more repulsive than the KL potential.

The propagation of kaons is described by Hamiltonian equations of motion
with the Hamilton function given by Eq.(\ref{omega}).
We also use the in-medium thresholds of the cross sections for kaon
production at SIS energies \cite{Fuchs01}.

\section{Numerical results}

\begin{figure}[htb]
\vspace*{-0.5cm}
                  \insertplot{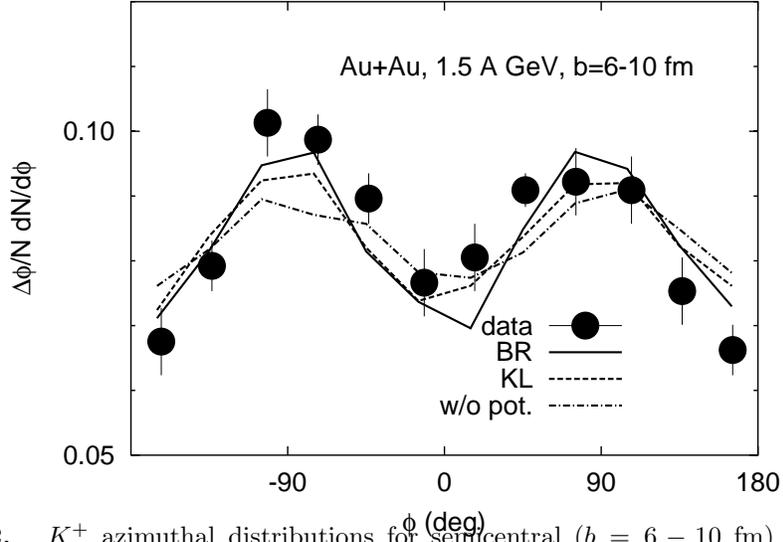}
\vspace*{-1.0cm}
\caption[]{
$K^+$ azimuthal distributions for semicentral ($b=6-10$ fm) Au+Au collisions 
at 1.5 A GeV. Kaons are selected in the rapidity range $|Y^{(0)}|<0.4$ and in 
the transverse momentum range $p_t=0.2-0.8$ GeV/c.
Data are from Ref. \cite{Uhlig05}.}
\label{fig:azdst_au148au}
\end{figure}

Fig.~\ref{fig:azdst_au148au} shows the 
azimuthal distributions of $K^+$'s at midrapidity from semicentral Au+Au 
collisions at 1.5 A GeV. The experimental data reveal a pronounced 
out-of-plane emission of $K^+$'s. We see, however, that the squeeze-out 
signal is clearly too weak in the calculation without
kaon potential.  The KL parametrization also produces not enough anisotropy.
The best description of the data is reached  in the calculation with the  
BR parametrization of the kaon mean field. The mechanism of the kaon 
squeeze-out enhancement is a dynamical focusing by the repulsive mean field 
\cite{LM05}. This seems to be different from the nucleon squeeze-out, which
is mostly due to shadowing by spectators \cite{com1}.

\begin{figure}[htb]
\vspace*{-0.5cm}
                   \insertplot{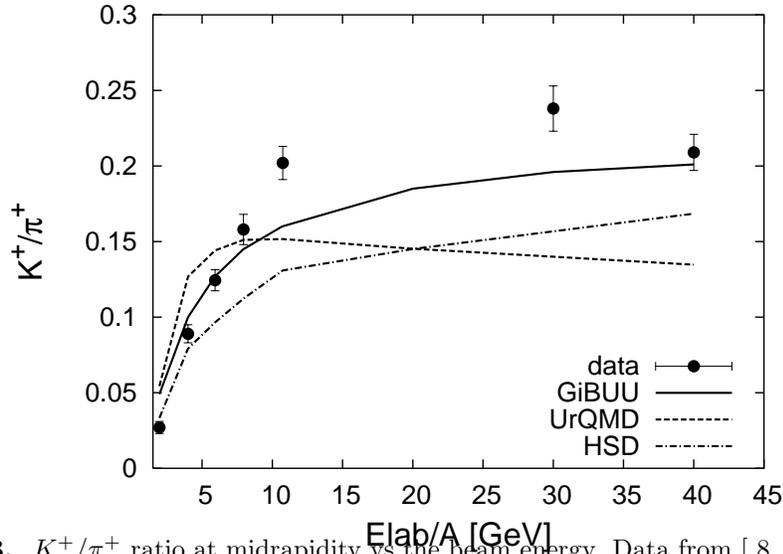}
\vspace*{-1.0cm}
\caption[]{ $K^+/\pi^+$ ratio at midrapidity vs the beam energy. 
Data from \cite{Ahle00,Afan02,Friese04}. }
\label{fig:ratio} 
\end{figure}

Fig.~\ref{fig:ratio} shows the $K^+/\pi^+$ ratio at midrapidity for
the central Au+Au (2-10 A GeV) and Pb+Pb (30 and 40 A GeV) collisions.
The GiBUU results are compared with the UrQMD and HSD calculations
from Ref. \cite{Weber03}. Although the GiBUU model gives a better 
description of the data, all calculations underpredict the ratio
at 10-30 A GeV due to overstimated pion production (c.f. 
\cite{WLM05,Weber03}). The GiBUU model produces somewhat more strangeness 
than the other models due to additional meson-meson channels for $K \bar K$ 
production.

\begin{figure}[htb]
\vspace*{-0.5cm}
                   \insertplot{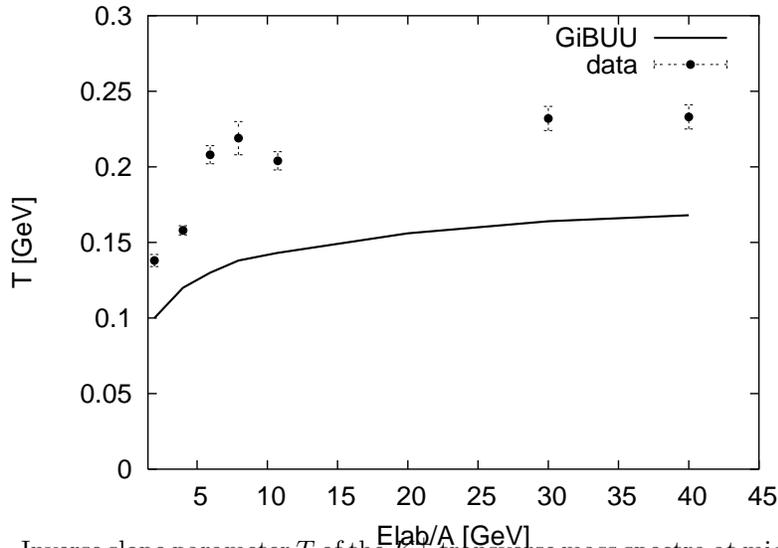}
\vspace*{-1.0cm}
\caption[]{ Inverse slope parameter $T$ of the $K^+$ transverse mass 
spectra at midrapidity obtained by a fit: 
$d^2\sigma/(2\pi m_\perp dm_\perp dy) = a \exp\{-m_\perp/T\}$. 
Data from \cite{Ahle00_1,Afan02,Friese04}.}
\label{fig:slope}
\end{figure}

While the kaon multiplicities are described rather well, the slope parameters of the
kaon transverse mass spectra are underpredicted by our model (Fig.~\ref{fig:slope}).

To understand the reason of the discrepancies above and to do a step towards
model improvement, we estimate the role of many-body collisions in
dense nuclear medium. In a central Au+Au collision at 20 A GeV the maximum
baryon density reached in a central 1 fm$^3$ cell is 
$\rho_B \simeq 10 \rho_0 = 1.6$ fm$^{-3}$. The gas parameter, i.e. the
number of particles inside the interaction volume characterizing a
two-body collision is
\begin{equation}
      \gamma_{gas}=(\sigma/\pi)^{3/2} \rho_B \simeq 2~,
\end{equation}
where $\sigma \simeq 40$ mbarn is the total baryon-baryon cross section.
Neglecting relativistic effects, one can conclude that the applicability 
condition of the Boltzmann equation, i.e. of the binary collision 
approximation, is violated (see also \cite{Mrow85}), since $\gamma_{gas} > 1$.
Relativistic effects, such as the Lorentz contraction of the interaction
volume along the collision axis, favour binary collisions at the initial
nonequilibrium stage, but quickly loose their importance at the high-density
equilibrated stage.

\section{Conclusions}

To summarize, we have performed the transport GiBUU calculations of kaon 
and pion production at 1-40 A GeV. We have found that --- at SIS energies --- 
the kaon potential is needed to describe the out-of-plane squeeze-out of kaons.
The BR parametrization ($U_K(\rho_0) \simeq 30$ MeV) is favoured.

At AGS --- lower SPS energies, standard GiBUU gives overall agreement with 
HSD and UrQMD on $\pi^+$ and $K^+$ multiplicities. Data on $\pi^+$ 
multiplicity are overestimated, $K^+$ multiplicity is well described. Our 
model produces too soft kaon $m_t$-spectra, also reported for UrQMD 
and HSD in \cite{Brat04}. We believe that the major problem with the standard 
transport calculations is neglecting the many-body collisions which are 
increasingly important at high densities.


\begin{thebibliography}{99}
\bibitem{com} At the AGS-SPS energies the mean field potentials were 
              switched off.
\bibitem{com1} In \cite{Dani00} the nucleon squeeze-out at 
               $E_{lab}=0.4-1.0$ A GeV has also been related to
the dynamical focusing. But another study \cite{LCGM00} has demonstrated
that the nucleon out-of-plane enhancement is mostly due to shadowing.
\end{thebibliography}
\end{document}